\documentclass[letterpaper, 10 pt, conference]{ieeeconf}
\usepackage{graphicx}      % include this line if your document contains figures
\usepackage{amssymb, amsfonts, amsmath, }
\usepackage{bbm}
\usepackage{epsfig, latexsym, amsfonts, amssymb, graphicx}
\usepackage{tikz} %add back

\newtheorem{theorem}{Theorem}
\newtheorem{lemma}[theorem]{Lemma}
\newtheorem{corollary}[theorem]{Corollary}
\newtheorem{remark}[theorem]{Remark}
\newtheorem{definition}[theorem]{Definition}

\title{\LARGE \bf
A Graph Theoretic Characterization of Perfect Attackability and Detection in Distributed Control Systems
}

\author{Sean Weerakkody~~~~Xiaofei Liu~~~~Sang H. Son~~~~Bruno Sinopoli% <-this % stops a space
\thanks{S. Weerakkody, X. Liu, and B. Sinopoli are with the Department of Electrical and Computer Engineering, 
			Carnegie Mellon University, Pittsburgh, PA, USA 15213. Email: {\tt\small sweerakk@andrew.cmu.edu, xiaofeil@andrew.cmu.edu, brunos@ece.cmu.edu}}
\thanks{S. Weerakkody is supported in part by the Department of Defense (DoD) through the National Defense Science \& Engineering Graduate Fellowship (NDSEG) Program.}}

\begin{document}
\maketitle
\begin{abstract}
This paper is concerned with the analysis and design of secure Distributed Control Systems in the face of integrity attacks on sensors and controllers by external attackers or insiders. In general a DCS consists of many heterogenous components and agents including sensors, actuators, controllers. Due to its distributed nature, some agents may start misbehaving to disrupt the system. This paper first reviews necessary and sufficient conditions for deterministic detection of integrity attacks carried out by any number of malicious agents, based on the concept of left invertibility of structural control systems. It then develops a notion equivalent to structural left invertibility in terms of vertex separators of a graph. This tool is then leveraged to design minimal communication networks for DCSs, which ensure that an adversary cannot generate undetectable attacks. Numerical examples are included to illustrate these results.

%The design and analysis of distributed control systems to prevent attacks is considered in this paper. In general a distributed control system consists of many heterogenous components and agents including sensors, actuators, controllers, and intrusion detection systems. This paper considers the scenario where an adversary attacks a subset of a system's sensors and actuators. Previous results have found that if a system is not left invertible an adversary can bias the state without generating an alarm from the intrusion detection system. This condition is also characterized graphically through the notion of structural left invertibility. In this article, we develop a notion equivalent to structural left invertibility in terms of vertex separators. This tool is then leveraged to design minimal communication networks for distributed control systems, which ensure that an adversary can not generate undetectable attacks. Numerical examples are included to illustrate these results.
\end{abstract}
\section{Introduction}
Distributed Control Systems (DCS) have become increasingly important in today's world. Unlike centralized control systems which use a single controller to perform actions, DCS may contain multiple controllers, possibly connected by communication networks, that manipulate their local environment to achieve a global purpose, based on local information. Due to advances in sensing, computing, and communication along with increasing reliance on large scale systems, DCS are found in a variety of applications. These include the smart grid \cite{Blaabjerg2006},\cite{Amin2005}, water management systems, process plants, sensor networks \cite{Sinopoli03}, formation control of autonomous vehicles \cite{Olfati2002}, \cite{Ren2008}, and average consensus \cite{Olfati2007}. 

Often reliant on off-the-shelf networking and directly linked to our critical infrastructures, DCS need to be secured against malicious attacks by developing tools to prevent, detect, and recover from intrusions \cite{Cardenas:2008ke}.  Precedent and motive to attack control systems have been documented. One significant example is Stuxnet, an attack on a Supervisory Control and Data Acquisition system in an uranium enrichment process plant in Iran. The attack infected local system controllers and fed false data to the intrusion detection system \cite{Langner2013}. An additional example is the Maroochy Shire incident where a disgruntled employee performed an attack on a sewage control system \cite{Slay2008}. 

Recent research efforts have endeavored to characterize and provide tools to defend against attacks on control systems. For instance, \cite{Mo2014} and \cite{Mo2009R} consider replay attacks on control systems and propose using a watermarked input to detect such attacks. In addition, \cite{liu2009} and \cite{henrik2010} consider attacks on the electricity grid.  \cite{liu2009}  provides conditions under which an adversary with knowledge of the systems structure can cause errors in state estimation. Moreover, \cite{henrik2010} proposes multiple security indices for sensors which allow a system operator to identify sparse attacks and add additional resources such as redundant sensors or encryption schemes as needed. 

Additionally, \cite{PasqualettiConsensusUnreliable} and \cite{Sundaram_function} consider attacks on DCS where the agents aimed to perform consensus. In particular, \cite{Sundaram_function}  proves that a given node calculating an arbitrary function can tolerate up to $f$ faulty agents if and only if there exist at least $2f +1$ vertex disjoint paths to any non-neighboring node. Additionally, \cite{PasqualettiConsensusUnreliable} characterizes attack identifiability and detectability graphically in terms of connectivity and algebraically using left invertibility.  \cite{wirelesscontrol} considers a DCS with a subset of malicious nodes.  The authors consider the design of an intrusion detection system, which can recover true system outputs as well as identify malicious nodes. \cite{Fawzi_secure_cont_est} considers the problems of robust control and estimation in the presence of attacks and proposes a practical decoder to perform detection.

\cite{Cam2014}  characterizes DCS that are perfectly attackable, where an adversary's actions have no effect on the output response, and provide algebraic and topological attackability conditions. \cite{PasqualettiAttack} introduces the notion of structural left invertibility to graphically characterize attack detectability. Namely the authors show that if a graphical realization of a control system is not structurally left invertible, the system is perfectly attackable, while, if the graph is structurally left invertible, almost all realizations of the system will not be perfectly attackable. 

In this article, we extend the analysis work of \cite{Cam2014} and \cite{PasqualettiAttack} to the design of minimal communication topologies capable of guaranteeing detection of integrity attacks on DCS. We consider the setting of a DCS where no more than $p$ agents may be compromised, thus corrupting sensor data and modifying control policies with the goal of disrupting system's operations in stealthy manner, i.e. without being detected. We assume a detector knows which control policies each agent is supposed to enforce; we also assume it collects sensor information from a subsets of the agents. With these assumptions we are able to characterize systems which are not perfectly attackable regardless of the adversary's attack policy, as a function of sensor placements and communication topology among the agents, based on the concept of left invertibility of structural control systems~\cite{Cam2014, PasqualettiAttack}. Our main contribution consists in providing design principles to guarantee that any attack will be revealed by the detector. To do so we first obtain an equivalent notion of structural left invertibility through the use of vertex separators. This notion allows to pose the task of designing the minimal communication topologies as optimization problems. In the first problem, given a fixed number of observers larger than $p$, we find the minimum number of communication links that can guarantee perfect detectability. We show how the  problem of jointly minimizing the number of sensors and communication links strictly depends upon the cost of sensing and communicating for the general case. If sensing is more expensive, we show that the optimal choice is to set the number of sensors to equal the number of malicious nodes $p$ and minimize over the communication topology, as in the previous case. If sensing is cheaper than communicating we prove that the optimal solutions is to deploy as many sensors as the number of states. In the case where only controllers, but not observers, can be attacked, we formulate an optimization problem to compute the solution.
% but still easy to compute. 
%In the unconstrained case, any node can talk any other node in the network, the minimum topology can be recovered along with theit 
 %As a consequence optimal designs can be derived in closed form. A non trivial
%Additionally, we consider an optimization where we aim to maximize the sparsity of the network while ensuring that the system is not perfectly attackable. Closed form results are obtained. 

The rest of the paper is formulated as follows. In section II, we provide descriptions of our control system both algebraically and graphically. In section III, we introduce an attack model, define perfectly attackable systems and revisit structural left invertibility. In section IV, we provide graphical conditions for a system to be perfectly attackable for some feasible attack input using vertex separators. In section V, we formulate an optimization problem to minimize the amount of communication in the system while ensuring that the network is robust to perfect attacks. In section VI, we include a numerical example and in section VII, we conclude the paper.
\section{System Model}
In this section, we introduce our model of a DCS. We assume that there are $n$ agents, $x_1, \cdots, x_n$ communicating with each other, and that they are observed by $m$ observers, $y_1, \cdots, y_m$ where $m \le n$. For simplicity we let $\mathcal{X} \triangleq \{x_1, \cdots, x_n\}$ and $\mathcal{Y} \triangleq \{y_1, \cdots, y_m\}$. We model their interactions with a directed graph $\mathcal{G} = (\mathcal{V},\mathcal{E})$ where $\mathcal{V} = \mathcal{X} \cup \mathcal{Y}$ is the set of agents and observers. Here $\mathcal{E} \subset \mathcal{V} \times \mathcal{V}$ represents communication between agents or observation of an agent by an observer. That is, if $(x_i,x_j) \in \mathcal{E}$, then $x_i$ may communicate with $x_j$. Furthermore, if $(x_j, y_i) \in \mathcal{E}$, then observer $y_i$ observes node $x_j$. 
%We assume the graph has no multiple edges.

The set of incoming neighbors to a node $a$ is given as 
\begin{equation}
N_a^I = \{b \in \mathcal{V}:~(b,a) \in \mathcal{E}\}.
\end{equation}
The in-degree of node $a$, denoted as $d_a^I$ is given by $d_a^I = |N_a^I|$. Similarly, the set of outgoing neighbors is defined as
\begin{equation}
N_a^O = \{b \in \mathcal{V}:~(a,b) \in \mathcal{E}\}.
\end{equation}
Here the out-degree $d_a^0$ of node $a$ is given by $d_a^O = |N_a^O|$. We assume that every node in $\mathcal{X}$ has a self loop.

Suppose $A \subset \mathcal{V}$ and $B \subset \mathcal{V}$. A path from $A$ to $B$ is a sequence of vertices $a_1,a_2,\cdots,a_l$ where $a_1 \in A$, $a_l \in B$, and $(a_j,a_{j+1}) \in \mathcal{E}$ for $1 \le j \le l-1$. A simple path contains no repeated vertices. Two paths are disjoint if they have no common vertices and two paths are internally disjoint if they have no common vertices except for possibly the starting and ending vertices. In general $l$ paths are disjoint if any pair of paths are disjoint. A set of $l$ disjoint and simple paths from a set $A$ to a set $B$ is called a $l$-linking from $A$ to $B$.

We assume that each agent in $\mathcal{X}$ is associated with a scalar state which is dependent on time. The state of a node $x_i$ at time $k$ is given by $x_i(k)$. We assume that the dynamics for each agent $x_i$ in $\mathcal{X}$ is given by
\begin{equation}
x_i(k+1) = a_{i,i}x_i(k) + u_i(k) + w_i(k).
\end{equation}
$w_i(k)$ is the process noise in the system and $u_i(k)$ is the input to the agent. Here we assume that $u_i(k)$ can be written as follows
\begin{equation}
u_i(k) = \sum_{j \in N_{x_i}^I, j \neq i} a_{i,j} x_j(k).
\end{equation}
Here we assume that at each time step, node $x_i$ receives the state from each of his incoming neighbors and computes a control law which is a linear function of its incoming neighbor's states. 
\begin{remark}
Although we assume that each input is a linear function of the state, in general we can consider an input of the form $u_i(k) = \sum_{j \in N_{x_i}^I, j \neq i} a_{i,j} x_j(k) + u_i^*(k)$, as long as the anomaly detection center knows $u_i^*(k)$ for all $x_i \in \mathcal{X}$ and for all $k$. Additionally the state $x_i(k)$ can refer to a physical quantity such as velocity, or can simply be a number associated with the node, for instance a value used for consensus.
\end{remark}

As mentioned earlier, a set of observers $\mathcal{Y}$ observes a subset of the agents. We assume each observer measures exactly one agent, and no two observers measure the same agent. Thus, if observer $y_i$ measures agent $x_j$, then we have
\begin{equation}
y_i(k) = x_j(k) 
\end{equation}
To simplify notation, we define
\begin{align*}
 x(k) &\triangleq \begin{bmatrix} x_1(k) & \cdots & x_n(k) \end{bmatrix}^T \in \mathbb{R}^n, \\
 w(k) &\triangleq \begin{bmatrix} w_1(k) & \cdots & w_n(k) \end{bmatrix}^T \in \mathbb{R}^n, \\
y(k) &\triangleq \begin{bmatrix} y_1(k) & \cdots & y_m(k) \end{bmatrix}^T \in \mathbb{R}^m. 
\end{align*}
Then, we have 
\begin{equation}
x(k+1) = Ax(k) + w(k), ~~ y(k) = Cx(k),
\end{equation}
where $A \triangleq [a_{i,j}]$ and the matrix $C$ is defined entrywise as
\begin{equation}
C_{ij} = \mathbf{1}_{(x_j,y_i) \in \mathcal{E}}.
\end{equation}
where $\mathbf{1}$ is the indicator function. The matrix $A$ is assumed to be stable and we assume $w(k) \sim \mathcal{N}(0,Q)$ is IID with $Q \ge 0$. A centralized detector is used to detect anomalies in the system. The centralized detector receives sensor measurements and uses a linear filter to perform estimation. We assume the centralized detector is aware of each agent's update rule, that is it knows $A$ as well as $C$. Furthermore, if an agent changes its update rule, it notifies the detector. The centralized detector uses the linear filter 
\begin{align}
\hat{x}(k+1) &= (A-KCA)\hat{x}(k) + Ky(k+1), 
\end{align}
to estimate the state, where $(A-KCA)$ is stable. Here $\hat{x}(k)$ is a state estimate of the state $x(k)$. A $\chi^2$ detector is used to detect the presence of abnormalities or an attack. In particular, an alarm is triggered if 
\begin{equation}
z(k)^T\mathcal{P}^{-1}z(k) > \eta
\end{equation}
where $z(k) = y(k) - CA\hat{x}(k-1)$ is the residue, $\mathcal{P}$ is the covariance of the residue, and $\eta$ is the threshold.
\section{Attack Model}
\subsection{Attack Description}
In this section we describe the attack model and define the concept of a perfectly attackable system. We assume that at time $0$ a subset of nodes in $\mathcal{V}$ is compromised. For instance, if an attacker is able to locally corrupt an agent, it likely will also have physical access to the sensor which measures it. The set of compromised nodes are denoted by $F \subset \mathcal{V}$. The set of compromised nodes are unknown to the centralized detector and can be comprised of both agents and observers. However, it is known that $|F| \le p$. Suppose that $F = \{x_{i_1}, \cdots, x_{i_l}, y_{i_{l+1}}, \cdots y_{i_{p^\prime}}\}$ where $p^\prime \le p$. In this case, the defender must choose how many compromised nodes it wishes to tolerate.
The set of all feasible compromised nodes is given by
\begin{equation}
\mathcal{F}_{xy} = \{F \subset \mathcal{V}, |F| \le p\}.
\end{equation}
We will later also consider a set of attacks $\mathcal{F}_{x}$, where an attack is restricted to agents. That is, no observers are attacked.
\begin{equation}
\mathcal{F}_{x} = \{F \subset \mathcal{X}, |F| \le p\}.
\end{equation} 
 We define the set of attack inputs to be $\mathcal{U} \triangleq \{u_1^a, \cdots u_{p^\prime}^a\}$. We model our system with a directed graph $\mathcal{G}^a = (\mathcal{V}^a,\mathcal{E}^a)$ where $\mathcal{V}^a = \mathcal{V} \cup \mathcal{U}$ and $\mathcal{E}^a = \mathcal{E} \cup \mathcal{E}_{\mathcal{U},\mathcal{X}} \cup \mathcal{E}_{\mathcal{U},\mathcal{Y}} $ where 
\begin{equation}
\mathcal{E}_{\mathcal{U},\mathcal{X}} = \{ (u_1^a,x_{i_1}), \cdots, (u_{l}^a,x_{i_l}) \} ,
\end{equation}
\begin{equation}
\mathcal{E}_{\mathcal{U},\mathcal{Y}} = \{(u_{l+1}^a,y_{i_{l+1}}), \cdots (u_{p^\prime}^a, y_{i_{p^\prime}})\}.
\end{equation}
As a result, we now add additional attack inputs nodes to our system digraph to represent compromised sensors or agents. Thus, if an agent or sensor node is compromised it now has an incoming edge from an attack input.

We let $x^a(k)$ denote the state of the compromised system and $y^a(k)$ denote the output. If an agent $x_i$ is compromised by an input $u_l^a$ we assume its update rule follows
\begin{equation}
x_i^a(k+1) = a_{i,i}x_i^a(k) + \sum_{j \in N_{x_i}^I, j \neq i} a_{i,j} x_j^a(k) + u_{l}^a(k) + w_{i}(k).
\end{equation}
If an agent $x_i$ is not compromised the original update rule follows
\begin{equation}
x_i^a(k+1) = a_{i,i}x_i^a(k) + \sum_{j \in N_{x_i}^I, j \neq i} a_{i,j} x_j^a(k)  + w_{i}(k).
\end{equation}
If an observer $y_i$ measuring $x_j$ is compromised by input $u_l^a$, then its measurement is given by
\begin{equation}
y_i^a(k) = x_j^a(k) + u_{l}^a(k) + v_i(k).
\end{equation}
Thus, compromised nodes have additive attacks. 
Finally, to simplify notation, we define $B^a \in \mathbb{R}^{n \times p^\prime}$ and $D^a \in \mathbb{R}^{m \times p^\prime}$ entrywise as 
\begin{equation}
B^a_{ij} = \mathbf{1}_{(u_j^a,x_i) \in \mathcal{E}_{\mathcal{U},\mathcal{X}}},~~ D^a_{ij} = \mathbf{1}_{(u_j^a,y_i) \in \mathcal{E}_{\mathcal{U}, \mathcal{Y}}}.
\end{equation}
Let $u^a(k) = \begin{bmatrix} u_1^a(k) & \cdots & u_{p^\prime}^a(k) \end{bmatrix}$. Thus, when under attack, the DCS has dynamics given by 
\begin{align}
x^a (k+1) &= A x^a(k) + B^a u^a(k) + w(k), \\
y^a(k) &= C x^a(k) + D^au^a(k) + v(k).
\end{align}
Moreover, under attack the estimator has dynamics given by
\begin{align}
\hat{x}^a(k+1) &= (A-KCA)\hat{x}^a(k) + Ky^a(k+1), \\
z^a(k) &= y^a(k) - CA\hat{x}^a(k-1). 
\end{align}

Now consider the difference between the compromised system and the system operating normally. We define the following variables
\begin{equation}
\Delta x(k) \stackrel {\Delta}{=} x(k) - x^a(k), ~~~~\Delta \hat{x}(k) \stackrel {\Delta}{=} \hat x(k) - \hat x^a(k),
\end{equation}
\begin{equation}
\Delta y(k) \stackrel {\Delta}{=} y(k) - y^a(k),~~~~\Delta {z}(k) \stackrel {\Delta}{=} z(k) - z^a(k).
\end{equation}

The goal of an adversary in a DCS is to affect the state of the distributed system without raising an alarm in the centralized detector. To characterize this, we introduce the following definition.

\begin{definition} 
An attack is perfect if $\Delta z(k) = 0$ for all $k \ge 0$ and $u^a(k) \neq 0$ for some $k \ge 0$. A system $(A,B^a,C,D^a)$ is perfectly attackable if there exists a perfect attack.
\end{definition}
\begin{remark}
If an attack is perfect, the residues of the system operating normally and the residues under attack are the same. Thus, the centralized detector can not distinguish an attack from normal operation. However, while under attack the adversary is able to bias the state away from normal operation. Note, that in practice $\Delta z(k)$ need not be $0$ to avoid detection. \cite{Cam2014} for instance considers the notion of a nearly perfect attack where an adversary can destabilize a system with bounded effect on the residues. In this paper, however, we restrict our attention to perfect attacks.
\end{remark}
\subsection{Conditions for Perfect Attackability}
We now briefly review both algebraic and graphical conditions which allow for a system to be perfectly attackable. To begin we introduce left invertibility.
\begin{definition}
We define a system $(A,B^a,C,D^a)$ to be left invertible if for the following system
\begin{equation}
x(k+1) = Ax(k) + B^au(k), ~~y(k) = Cx(k) + D^au(k),
\end{equation}
with initial condition $x(0) = 0$, $y(k) = 0$ for all $k$ implies that $u(k) = 0$ for all $k$.
\end{definition}
 It can be shown that the left invertibility of a system is necessary and sufficient for a system to be perfectly attackable. In particular we have the following theorem from \cite{Cam2014}.

\begin{theorem}
The following statements are equivalent.
\begin{enumerate}
\item There exists a sequence of inputs $u^a(k) \neq 0$ such that $\Delta z(k) = 0$ for all $k$.
\item There exists a sequence of inputs $u^a(k) \neq 0$ such that $\Delta y(k) = 0$ for all $k$.
\item $(A,B^a,C,D^a)$ is not left invertible.
\item The transfer function $C(zI-A)^{-1}B^a + D^a$ has normal rank less than $p^\prime$.
\end{enumerate}
\end{theorem}
The last statement gives us means to algebraically verify if a system is left invertible. We now look to graphically characterize when a system is perfectly attackable. To do this we consider structural linear systems \cite{Lin1974}. Here we associate the graph $\mathcal{G}^a$ with a tuple of structural matrices $([A], [B^a], [C], [D^a])$. We observe that $\mathcal{E}^a = \mathcal{E}_{\mathcal{X},\mathcal{X}} \cup \mathcal{E}_{\mathcal{U},\mathcal{X}}  \cup \mathcal{E}_{\mathcal{X},\mathcal{Y}}   \cup \mathcal{E}_{\mathcal{U},\mathcal{Y}}$ where $\mathcal{E}_{\mathcal{X},\mathcal{X}} = \{(x_i,x_j): [A]_{ji} \neq 0\}$, $\mathcal{E}_{\mathcal{U},\mathcal{X}} = \{(u_i,x_j): [B^a]_{ji} \neq 0\}$, $\mathcal{E}_{\mathcal{X},\mathcal{Y}} = \{(x_i,y_j): [C]_{ji} \neq 0\}$, and $\mathcal{E}_{\mathcal{U},\mathcal{Y}} = \{(u_i,y_j): [D]^a_{ji} \neq 0\}$. Also $[A]_{ij} \neq 0$ means that $A_{ij}$ is a free parameter while $[A]_{ij} = 0$ implies that $A_{ij}$ is fixed to be 0.

We would like to use structural systems to obtain a graphical characterization of left invertibility. In particular, we have the following definition
\begin{definition}
The structural system $([A], [B^a], [C], [D^a])$ is structurally left invertible if every admissible realization of $(A,B^a,C,D^a)$ is left invertible with the exception of a set of measure $0$.
\end{definition}
We note here that if $([A], [B^a], [C], [D^a])$ is not structurally left invertible, then every admissible realization of $(A,B^a,C,D^a)$ is also not left invertible. Thus, if we can ensure that a graphical realization of a system is structurally left invertible, then almost all numerical realizations of that system will be left invertible and as a result, not perfectly attackable. We have the follow result that characterizes structural left invertibility from \cite{PasqualettiAttack}.
\begin{theorem}
The system $([A],[B^a],[C],[D^a])$ is structurally left invertible if and only if there exists a linking of size $|\mathcal{U}|$ from $\mathcal{U}$ to $\mathcal{Y}$.
\end{theorem}
From this result we see that a necessary condition for a system to be left invertible is that there exist more sensors than attack inputs. As a result, we have
\begin{corollary}
The system $([A],[B^a],[C],[D^a])$ is structurally left invertible only if $m \ge p^\prime$.
\end{corollary}

\section{Vertex Separators and Structural Left Invertibility}
In the previous section we showed that for almost all realizations of a system $(A,B^a,C,D^a)$, structural left invertibility is equivalent to a system not being perfectly attackable. In this section we obtain an equivalent characterization of structural left invertibility using the notion of vertex separators. In particular, we can use vertex separators to characterize systems that are structurally left invertible for all feasible attacks.
We begin by defining vertex separators.
\begin{definition}
 Given a  graph $\mathcal{G^*} = (\mathcal{V^*},\mathcal{E^*})$, a vertex separator $S \subset \mathcal{V}^*$ of nonadjacent vertices $(a,b)$ is a subset of vertices whose removal  removes all paths from $a$ to $b$.
 \end{definition}
 We now consider how vertex separators can be used to characterize structural left invertibility for a system with attacks on both nodes and sensors.
%A vertex separator $S$ is minimal vertex separator of $(a,b)$ if there is no $S^\prime \subset S$ which is also a vertex separator of $(a,b)$.
\begin{theorem}
Consider system $([A],[B^a],[C],[D^a])$ with feasible attack policy $\mathcal{F}_{xy}$ and corresponding graph realization $\mathcal{G}^a$ where $m \ge p$ dedicated sensors are assigned to measure a portion of the state. Here a dedicated sensor means that the sensor measures exactly one state. Suppose we obtain the graph $\mathcal{G}^\prime = (\mathcal{X} \cup \mathcal{Y} \cup o, E^\prime)$ by adding an additional node $o$ to $\mathcal{G}^a$ with directed edges from $\mathcal{Y}$ to $o$. The inputs $\mathcal{U}$ are also removed. Then $([A],[B^a],[C],[D^a])$ is structurally left invertible for all feasible attack policies $\mathcal{F}_{xy}$ if and only if for each node $x_i \in \mathcal{X}$, all vertex separators $S_i$ of $(x_i,o)$ in $\mathcal{G}^\prime$ satisfy $|S_i| \ge p$.  \label{MengerCor}
\end{theorem}
\begin{proof}
\textbf{ $\Rightarrow$}:
Suppose there exists a vertex separator $S_1 = \{x_2, \cdots, x_j, y_{j+1}, \cdots,  y_{l+1}\}$ of  $(x_1,o)$  in $\mathcal{G}^\prime$  with $|S_1| = l < p$. We observe, by definition $o \notin S_1$, so that $S_1 \subset \mathcal{X} \cup \mathcal{Y}$. Suppose an adversary inserts dedicated inputs $u_i$ to $x_i$ for $1 \le i \le j$ and dedicated inputs $u_i$ to $y_i$ for $j+1 \le i \le l+1$. Here a dedicated input is fed to exactly one state. Such an attack is feasible because $l + 1 \le p$. Any path from $u_i$ to $o$ must contain $x_i$ for $1 \le i \le j$ or $y_i$ for $j+1 \le i \le l+1$ . Let $P_j \subset u_j \cup \mathcal{X} \cup \mathcal{Y}$ denote the set of vertices in a path from $u_j$ to $o$ excluding $o$. Consequently $S_1 \subset P_2 \cup P_3 \cup \cdots \cup P_l$. 

A necessary condition for there to be a $p$-linking is the existence of a path $P_1$ such that $P_1 \cap ( P_2 \cup P_3 \cup \cdots \cup P_l) = \emptyset$. As a result, for there to be a $p$-linking $P_1 \cap S_1 = \emptyset$. However, by definition of a vertex separator any path in $\mathcal{G}^\prime$ from $x_1$ to $o$ must contain an element of $S_1$ and thus any path from $u_1$ to $o$ must also contain an element in $S_1$. Consequently, $P_1 \cap S_1 \neq \emptyset$ and there is no $p$-linking. Thus, the system is not structurally left invertible.

\textbf{$\Leftarrow$}:
Suppose all vertex separators $S_i$ of $(x_i,o)$ in $\mathcal{G}^\prime$ satisfy $|S_i| \ge p$ where $x_i \in \mathcal{X}$. To begin we observe from Menger's Theorem \cite{Menger1927} that the minimum size of a vertex separator between $x_i$ and $o$ is equal to the maximum number of internally disjoint paths from $x_i$ to $o$. As a result, for each $x_i \in \mathcal{X}$ there exist at least $p$ internally disjoint paths from $x_i$ to $o$ which we can assume are simple by removing all cycles.

Now suppose WLOG an adversary implements a feasible attack policy with $p^\prime = p$ where he attacks $l \ge 0$ agents $x_1, \cdots, x_l$ and $p - l$ sensors $y_{l+1}, \cdots, y_p$. Let us now define a graph $\mathcal{G}^*$ obtained by adding an additional vertex $u$ to $\mathcal{G}^\prime$ with directed edges to all the attacked nodes. We wish to show that the minimum size of a vertex separator between $u$ and $o$ is at least $p$.  To do this, suppose there exists a vertex separator $S_u$ such that $|S_u| = k < p$. Since $|S_u| < p$, the number of attacked nodes, there must exist a node $z \in \{x_1, \cdots, x_l,y_{l+1}, \cdots, y_p\}$ such that $z\notin S_u$. Suppose $z$ is an observer. Then there exists a path $u, z, o$, even when nodes in the vertex separator are removed. If $z$ is an agent, there are at least $p$ internally disjoint paths from $z$ to $o$. Since $|S_u| < p$, at least one of these paths $z, P^*, o$ satisfies $P^* \cap S_u = \emptyset$. Thus, $u, z, P^*, o$ forms a simple path from $u$ to $o$, even when all the nodes in the vertex separator $S_u$ are removed. This contradicts the assumption that $|S_u| < p$. As a result, from Menger's Theorem, there exist at least $p$ internally disjoint and simple paths from $u$ to $o$ on $\mathcal{G}^*$.

If there are $p$ internally disjoint, simple paths from $u$ to $o$ on $\mathcal{G}^*$, there are $p$ disjoint and simple paths from $\mathcal{U}$ to $\mathcal{Y}$ on $\mathcal{G}^a$. As a result, there exists a $p$-linking and the system is structurally left invertible.
\end{proof}
We now extend these results to an attack restricted to just agents, $D^a = 0$.
\begin{corollary}
Consider structural system  $([A],[B^a],[C])$ with feasible attack policy $\mathcal{F}_{x}$ and corresponding graph realization $\mathcal{G}^a$ where $m \ge p$ dedicated sensors are assigned to measure a portion of the state. Suppose we obtain the graph $\mathcal{G}^\prime = (\mathcal{X} \cup o, \mathcal{E}^\prime)$ by collapsing all observer nodes to a single node $o$ and removing the inputs $\mathcal{U}$. Then $([A],[B^a],[C])$ is structurally left invertible for all feasible attack policies $\mathcal{F}_{x}$ if and only if for each unobserved $x_i \in \mathcal{X}$, all vertex separators $S_i$ of $(x_i,o)$ in $\mathcal{G}^\prime$ satisfy $|S_i| \ge p$. \label{VS ABC} 
\end{corollary}
\begin{proof}
    The proof of sufficiency follows an identical argument to that of Theorem \ref{MengerCor} with the exception that all vertex separators lie in $\mathcal{X}$. The necessary argument also follows directly from Theorem \ref{MengerCor} where we replace references to all nodes in $\mathcal{X}$ with references to unobserved nodes $x_i$ and replace references to sensor nodes $y_i \in \mathcal{Y}$ with references to observed nodes.
\end{proof}
\begin{remark}
We note that Pasqualetti et. al. \cite{PasqualettiConsensusUnreliable} arrive at necessary conditions for a system to be structurally left invertible using the connectivity of the graph. However, they only consider connected graphs. Here, through the use of vertex separators we arrived at a both sufficient and necessary condition for structural left invertibility for all feasible attacks. This result furthermore illustrates that the digraph need not be connected in order to ensure the system is robust to perfect attacks.
\end{remark}

We note that a minimum vertex separator, a vertex separator with the fewest number of nodes, can be computed  within a poly-logarithmic factor of $M(n+p+m)$, where $M(n+p+m)$ denotes the number of arithmetic operations for multiplying two matrices in $\mathbb{R}^{(n+p+m) \times (n+p+m)}$ \cite{Cheriyan1997} Note here that $n+p+m$ is the number of vertices in $\mathcal{G}^a$. In the next section we will show how vertex separators can be used to design optimal networks.

\section{Minimal Design of Structurally Left Invertible Systems}
 %In large scale systems, many components are battery operated with limited power supply. Here communication to other agents uses a significant portion of a node's power. 
 In this section, we aim to minimize the number of communication links in a DCS while simultaneously achieving a system robust to perfect attacks from a feasible attack policy. To begin we include the following lemma.
\begin{lemma}
 $([A],[B^a],[C],[D^a])$ is structurally left invertible for all possible attack configurations $\mathcal{F}_{xy}$ only if the out-degree of each node $x_k \in \mathcal{X}$ in $G^a$ satisfies $d_{x_k}^{O} \ge {p+1}$. \label{NECLEMMA}
\end{lemma}
\begin{proof}
From Theorem \ref{MengerCor}, we only need to find a vertex separator $S_i$ of $(x_i,o)$ in $\mathcal{G}^\prime$ such that $|S_i| < p$ for some $x_i \in \mathcal{X}$. We choose $x_i$ such that $d_{x_i}^{O} < {p+1}$. We argue that a vertex separator of $(x_i,o)$ is the set of $x_i$'s outgoing neighbors $N_{x_i}^O \setminus x_i$. Here $|S_i| \le p-1$. If we remove all the outgoing neighbors of $x_i$, there is no path from $x_i$ to $o$ and the result holds.
\end{proof}

For a given number of observers $m$ and attackers $p \le m$, we wish to solve the following problem. \\
\textbf{Problem 1}
\begin{align*}
 & \underset{[A],[C]}{\min}  \| A \|_0 \\
 & \mbox{s.t. } ([A],[B^a],[C],[D^a]) \mbox{ is structurally left invertible} \\ 
 &\mbox{for all feasible attacks , i.e for all }F \in  \mathcal{F}_{xy}.
\end{align*}
Here to reduce the amount of communication, we aim to minimize the number of connections in the system. However, to preserve robustness, we ensure that the system is structurally left invertible for all feasible attacks.
\begin{theorem}
The optimal solution to problem 1 is $\|A\|_0^* =  mp + (n-m)(p+1)  = np + n - m$. \label{optimization}
\end{theorem}
\begin{proof}
We begin by showing that $np + n -m$ is a lower bound of the optimal solution $\|A\|_0^*$.  Without loss of generality, assume that $\{x_1,\cdots,x_m\}$ are observed nodes. We observe that 
\begin{align*}
\|A\|_0^*  &= \sum_{k=1}^m (d_{x_k}^{O*} - 1) + \sum_{k=m+1}^n d_{x_k}^{O*} , \\
  &\ge mp + (n-m)(p+1) . 
 \end{align*}
The first equality is obtained by noting that the number of nonzero entries for in each row $i$ of A is equal to $d_{x_k}^{O}$ if the node $x_i$ is unobserved and equal to $d_{x_k}^{O}-1$ if it is observed. The last inequality is obtained from the necessary conditions for structural left invertibility described in Lemma \ref{NECLEMMA}. Thus $np+n-m$ is a lower bound for $\|A\|_0^*$.

We now show that $np+n-m$ is an upper bound for $\|A\|_0^*$ by constructing  $([A],[C])$ so that $ ([A],[B^a],[C],[D^a])$ is structurally left invertible for all feasible attack policies $\mathcal{F}_{xy}$ and $\|A\|_0 = np +n - m$. A feasible configuration would be to select $m$ arbitrary nodes to observe. WLOG we assume that nodes $\{x_1,\cdots, x_m\}$ are observed so that there exists a directed edge from $x_j$ to $y_j$ for $j \in \{1,\cdots,m\}$. Next for $j \in \{1,\cdots,m\}$, we have $d_{x_j}^O = p+1$ and  $N_{x_j}^O \subset \{y_j, x_1, \cdots, x_m \}$. Thus, each observed node has $p+1$ outgoing edges, $1$ to its observer, $p-1$ edges to other observed nodes, and $1$ to itself. Finally, for $j \in \{m+1,\cdots, n\}$, we have $d_{x_j}^O = p+1$ and  $N_{x_j}^O \subset \{ x_1, \cdots,  x_m \}$. That is, each unobserved node has $p$ neighbors besides itself, all of which are observed nodes. Thus we have
\begin{align*}
\|A\|_0  &= \sum_{k=1}^m (d_{x_k}^{O} - 1) + \sum_{k=m+1}^n d_{x_k}^{O} , \\
  &= mp + (n-m)(p+1) .  
 \end{align*}

We now prove that the system is structurally left invertible for all feasible attack policies $\mathcal{F}_{xy}$ by showing that for each $x_i \in \mathcal{X}$, all vertex separators $S_i$ of $(x_i,o)$ in $\mathcal{G}^\prime$ from Theorem \ref{MengerCor} satisfy $|S_i| \ge p$. First suppose $x_i$ is an unobserved node and let $S_i$ be a vertex separator such that $|S_i| < p$. Since $|S_i|$ is less than $p$ and $x_i$ has outgoing edges to $p$ observed nodes, there exists observed node $x_j$ with observer $y_j$ such that $x_j, y_j \notin S_i$. Thus, even after removing nodes in $S_i$, the path $x_i, x_j, y_j, o$ exists. By contradiction $|S_i| \ge p$. 

Now suppose $x_i$ is an observed node and let $S_i$ be a vertex separator such that $|S_i| < p$. If $y_i$ is not in $S_i$, then $x_i, y_i, o$ forms a path from $x_i$ to $o$ which means $S_i$ is not a vertex separator. Now assume $S_i$ does contain $y_i$. Since $|S_i \setminus y_i| < p-1$, and $x_i$ has outgoing edges to $p-1$ observed nodes besides itself, there exists observed node $x_j \neq x_i$ with observer $y_j$ such that $x_j, y_j \notin S_i$. Thus, even after removing nodes in $S_i$, the path $x_i, x_j, y_j, o$ exists. By contradiction $|S_i| \ge p$. As a result, the system is structurally left invertible.
\end{proof}

Instead of fixing the number of sensors under consideration $m$, the number of sensors can be a design variable which is chosen concurrently with the network. The adjusted optimization problem is given as \\
\textbf{Problem 2}
\begin{align*}
 & \underset{[A],[C], m}{\min}  K_1\| A \|_0  + K_2 m\\
 & \mbox{s.t. } ([A],[B^a],[C],[D^a]) \mbox{ is structurally left invertible} \\ 
 &\mbox{for all feasible attacks , i.e for all }F \in \mathcal{F}_{xy}, \\
 & m \in \{p, p+1, \cdots, n\}.
\end{align*}
Note this problem is equivalent to 
\begin{align*}
 & \underset{m}{\min}  \left( \underset{[A],[C]}{\min}  K_1\| A \|_0(m)  + K_2 m \right),\\
 & = \underset{m}{\min}  ~ K_1n(p+1) + (K_2-K_1)m.
\end{align*}
Thus if $K_2 > K_1$ so that sensors are more costly than network links we simply take the minimum number of sensors, $m^* = p$. If $K_1 < K_2$, so that network links are more expensive, we take the maximum number of sensors, $m^* = n$.

Suppose that we instead consider the case where an adversary only attacks agents, that is the states $\mathcal{X}$, and does not directly manipulate sensors $\mathcal{Y}$. In this case, we are guaranteed that $D^a = 0$ and the feasible set of attack policies are described by $\mathcal{F}_x$. Since, the adversary has less surfaces in which he can attack, the optimal network is more sparse.
\begin{theorem}
Consider the following problem where we assume the attacker is limited to state attacks. \\
\textbf{Problem 3}
\begin{align*}
 & \underset{[A],[C]}{\min}  \| A \|_0 \\
 & \mbox{s.t. } ([A],[B^a],[C]) \mbox{ is structurally left invertible} \\ 
 &\mbox{for all feasible attacks , i.e for all }F \in \mathcal{F}_x.
\end{align*}
The optimal solution to the problem  is $\|A\|_0^* =  (n-m)p + n$. \label{optimization2}
\end{theorem}

The proof is similar to the proof given for  Theorem \ref{optimization} and is thus omitted. A minimal realization in such a proof has each unobserved node having exactly $p$ outgoing edges to observed nodes and one self loop. Each observed node would have one outgoing edge to an observer in addition to its self loop. We can again solve problem 2, for the case where $D^a = 0$. Here, for $K_2 > K_1 p$, we simply use the minimum number of sensors so $m^* = p$. If on the other hand, $K_2 < K_1 p$, we instead use the maximum number of sensors so $m^* = n$.
\begin{remark}
Theorems \ref{optimization} and \ref{optimization2} arrive at the minimum number of edges needed for there to exist a feasible structurally left invertible network. The proof of theorem \ref{optimization} additionally offers such a minimal realization. However, the general structure of an optimal network is currently unknown and must satisfy vertex separator conditions stated in Theorem \ref{MengerCor} and Corollary \ref{VS ABC}. Obtaining such a general structure is left as future work.
\end{remark}
%\documentclass[journal]{IEEEtran}
%\usepackage{graphicx}      % include this line if your document contains figures
%\usepackage[sort, numbers]{natbib}       % required for bibliography
%\usepackage{amssymb, amsfonts, amsmath, amsthm}
%\usepackage{bbm}
%\usepackage{epsfig, latexsym, amsfonts, amssymb, graphicx}
%\usepackage{tikz} %add back
%
%\newtheorem{theorem}{Theorem}
%\newtheorem{lemma}[theorem]{Lemma}
%\newtheorem{corollary}[theorem]{Corollary}
%\newtheorem{remark}[theorem]{Remark}
%\newtheorem{definition}[theorem]{Definition}
%\newtheorem{example}[theorem]{Example}
%%\input epsf
%
%\setcounter{page}{1}
%
%\begin{document}

\section{Numerical Example: Secure Platooning}\label{examples}

In this section, we provide an illustrative example, where the main results are used. More precisely, in this section, we use a simple platooning secure example to illustrate a minimal communication network robust to perfect attacks based on Theorem \ref{optimization}.  

\begin{figure}[htb]
\centering
\includegraphics[scale=0.4]{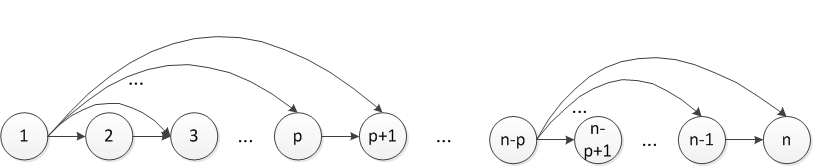}
\caption{Platooning Network, Self Loops Not Shown}
\label{fig:vehicles}
\end{figure}

The topic of secure platooning has been recently investigated by the research community \cite{Bruce2015}. Here, we consider a communication network of $n$ agents, with topology shown in Figure \ref{fig:vehicles}, which can be viewed as a simple platooning network. In this network, the vehicles are ordered in a line, and each vehicle can communicate with at most $p$ vehicles ahead of it. The control goal is to move vehicles at a constant speed while maintaining a minimum safety distance.

More formally, the communication network can be modeled as a digraph $\mathcal{G}=(\mathcal X \cup \mathcal Y,\mathcal E)$, where $\mathcal X=\{x_1,\cdots,x_n\}$ corresponds to the set of $n$ agents, $\mathcal Y=\{y_1,\cdots,y_m\}$  corresponds to the set of $m$ observers, and $\mathcal E=\mathcal E_{\mathcal X,\mathcal X}\cup \mathcal E_{\mathcal X,\mathcal Y}$ corresponds to the communication between agents, and communication between agents and observers, respectively. Taking attack configurations into consideration, the communication network can be represented as a quadruple $(A,B^a,C,D^a)$, where matrix $A$ is associated with $\mathcal E_{\mathcal X,\mathcal X}$, matrix $C$ is associated with $\mathcal E_{\mathcal X,\mathcal Y}$, and the number of observer nodes satisfies $m\geq p$. We first assume $D^a = 0$, so the adversary is limited to state attacks. In a platooning network, intuitively, we would expect the vehicles to communicate locally due to the difficulty and cost of distant communication. Ideally we would solve the following optimization problem
\begin{align}
 & \underset{[A],[C]}{\min}  \| A \|_0 \label{exprob} \\
 & \mbox{s.t. } ([A],[B^a],[C]) \mbox{ is structurally left invertible} \nonumber \\ 
 &\mbox{for all feasible attacks , i.e for all }F \in \mathcal{F}_x,\nonumber \\
 & \mathcal E_{\mathcal X,\mathcal X}\subset \{(x_i,x_{i+k})|1 \le i \le n, 0 \le k \le p,~i+k\leq n\}. \nonumber
\end{align}
We now prove that the optimal solution to \eqref{exprob} satisfies $\|A\|_0^* =  (n-m)p+n$. Since \eqref{exprob} enforces additional constraints compared to Problem 3, by Theorem \ref{optimization2} its optimal value satisfies $\| A \|_0^* \ge (n-m)p+n$. Now, we provide a feasible configuration pair $(A^f,C^f)$ so that the system is structurally left invertible for any feasible set of attackable nodes $F \in \mathcal{F}_x$ and $\|A^f\|_0=(n-m)p+n$. Now assume that
\begin{align*}
\mathcal E^f_{\mathcal X,\mathcal X} &=\{(x_i,x_{i+k})|i=1,\cdots,(n-m),k=0,\cdots,p\}, \\
& \cup \{(x_i,x_i)| n-m+1, \cdots, n\}, \\
\mathcal E^f_{\mathcal X,\mathcal Y} &=\{(x_{n-m+i},y_i)|i=1,\cdots,m\}.
\end{align*}
 Suppose we obtain the digraph $\mathcal{G}^\prime = (\mathcal{X}^\prime \cup o, \mathcal{E}^\prime)$ as described in Corollary \ref{VS ABC} . We are going to show that for each unobserved node $x_i \in \mathcal{X}$, $i=1,\cdots,n-m$, any minimum vertex separators $S_i$ of $(x_i,o)$ in $G^\prime$ satisfies $|S_i|=p$, and one minimum vertex separator is $\{x_{i+1},\cdots,x_{i+p}\}$. Each $x_i$ only has outgoing edges to $\{x_{i+1},\cdots,x_{i+p}\}$, hence $\{x_{i+1},\cdots,x_{i+p}\}$ is a vertex separator of $(x_i,o)$, according to the definition of a vertex separator. To show that $\{x_{i+1},\cdots,x_{i+p}\}$ is a minimum vertex separator, by Menger's theorem we can show that there exist $p$ vertex-disjoint paths from $x_{i+1},\cdots,x_{i+p}$ to $o$, respectively. Denote $d=(n-m+1)-(i+1)=\alpha p+\beta,~\alpha,\beta \in \mathbb{N}, \beta < p$, i.e., the distance between $x_{i+1}$, the nearest neighbor of $x_i$, and $x_{n-m+1}$, the nearest observed node. One collection of such vertex-disjoint paths is $\bigcup \limits_{k=1,\cdots,p} \{x_{i+k},x_{i+k+p}, \cdots, x_{i+k+(\alpha-1)p},$ $x_{i+k+\alpha p}, x_{n-m+k}\}$. As a result, $\{x_{i+1},\cdots,x_{i+p}\}$ is a minimum vertex separator. We can conclude that for each unobserved $x_i \in \{x_1,\cdots,x_{n-m}\}$, all vertex separators $S_i$ of $(x_i,o)$ in $G^\prime$ satisfy $|S_i| \ge p$. According to Corollary \ref{VS ABC}, $\{[A^f],[B^a],[C^f]\}$ is structurally left invertible.

Now, if $D^a\neq 0$, under the original constraint on $A$, it is impossible to design $(A,C)$ such that $([A],[B^a],[C],[D^a])$ is structurally left invertible for all feasible attacks, since for each node in $\{x_{n-p+1},\cdots,x_n\}$, the out-degree is less than $p$, which violates Lemma \ref{NECLEMMA}. In order to generate a feasible solution, we need to relax the restrictions, and let the agents in $\{x_{n-p+1},\cdots,x_n\}$ communicate with at least $p-1$ other observed agents. 
\begin{align}
 & \underset{[A],[C]}{\min}  \| A \|_0 \label{exprob2} \\
 & \mbox{s.t. } ([A],[B^a],[C]) \mbox{ is structurally left invertible}  \\ \nonumber 
 &\mbox{for all feasible attacks , i.e for all }F \in \mathcal{F}_{xy}, \nonumber \\
 & \mathcal E_{\mathcal X,\mathcal X}\subset \{(x_i,x_{i+k})|1 \le i \le n-p,~ 0 \le k \le p,~i+k\leq n\} \nonumber \\
 & \cup \{(x_i,x_j)| n-p+1 \le i \le n,~ x_j \mbox{ is observed}\}, \nonumber \\
 & N_{x_1}^O \ge p. \nonumber
\end{align}

In this case, the optimal solution to \eqref{exprob2} satisfies that $\|A\|_0^* = np + n -m$. According to Theorem \ref{optimization}, $\|A\|_0^* \geq np + n -m$. We now provide a feasible configuration $(A^{f'},C^{f'})$ such that $(A^{f'},B^a,C^{f'},D^a)$ is structurally left invertible for any feasible $(B^a,D^a)$, and $\|A^{f'}\|_0= np + n - m$. Specifically we let
\begin{align*}
 \mathcal E^{f'}_{\mathcal X,\mathcal X} &=\{(x_i,x_{i+k})|i=1,\cdots,(n-m),k=1,\cdots,p\}\cup \bar{\mathcal E} \\
 \mathcal E^{f'}_{\mathcal X,\mathcal Y} &=\{(x_{n-m+i},y_i)|i=1,\cdots,m\},
 \end{align*}
Here $\bar{\mathcal E}$ is defined so that we add outgoing edges from each observed agent to $p-1$ other observed agents and self loops. In this case, assume that the attacker attacks $p^*$ observers, where $p^*\leq p$, which is equivalent to removing $p^*$ observer nodes and $p^*$ attacks synchronously. The remaining communication network contains $m-p^*$ observed nodes and $n-m+p^*$ unobserved nodes, which can be represented through a structured system $([A^{f'}],[B^{a'}],[C'])$. We now obtain the digraph $\mathcal{G}^\prime = (\mathcal{X} \cup o, \mathcal{E}^\prime)$ as described in Corollary \ref{VS ABC}. We prove that for each unobserved $x_i \in \mathcal{X}$, all vertex separators $|S_i|$ of $(x_i,o)$ in $\mathcal{G}^\prime$ satisfy $|S_i| \ge p-p^*$. For each unobserved node $x_i\in\{x_1,\cdots,x_{n-m}\}$, the size of the minimum vertex separator is $p-p^*$. This is because there exist $p$ paths prior to removing observers and removing $p^*$ observers removes at most $p^*$ vertex-disjoint paths from $x_{i}$ to $o$. Now WLOG, assume that attacker attacks observer nodes $y_{1},\cdots,y_{p^*}$, then the set of nodes $\{x_{n-m+1}\cdots,x_{n-m+p^*}\}$ becomes unobserved. Note that each unobserved node $x_i\in\{x_{n-m+1}\cdots,x_{n-m+p^*}\}$ is still connected to at least $p-p^*$ observed nodes so the size of the minimum vertex separator $|S_i|$ of $(x_i,o)$ is $p-p^*$. Hence, according to Corollary \ref{VS ABC}, $([A^f],[B^a],[C^f],[D^a])$ is structurally left invertible.

\section{Conclusion}
In this article, we consider the setting of a DCS where a subset of up to $p$ agents and sensors may be compromise. We place a special focus on perfect attacks where an adversary can bias the system state without introducing a net effect on the output response. Previous work has shown that network topology determines the susceptibility of a DCS to a perfect attack based on the concept of structurally left invertible systems. Our main contributions include introducing the notion of vertex separators which allows us to graphically characterize systems which are resilient to perfect attacks regardless of an adversary's attack policy. 

We then use vertex separators to pose the problem of minimizing the number of communication links in our network. For a given number of sensors, we arrive at the minimum number of links which ensure resilience to perfect attacks as well as a feasible network realization. Furthermore, we considered the problem of jointly minimizing the network and the number of sensors. We determined that if sensing was more expensive than communicating it is optimal to minimize the number of sensors in a network while if communicating was more expensive than sensing it was optimal to observe the entire network and minimize communication. Future work includes applying these results to real large scale systems. In addition, we would like considering constraints on network communications when performing DCS design.  Finally we wish to investigate a larger class of attacks including the case where an adversary disrupts communication.

\bibliographystyle{IEEEtran}
\bibliography{Left_invert2}

\end{document}